\documentclass[twoside]{article}
\usepackage{fleqn,espcrc2}

\usepackage{graphicx}

\usepackage{latexsym}

\newcommand{\AmS}{{\protect\the\textfont2
  A\kern-.1667em\lower.5ex\hbox{M}\kern-.125emS}}

\newcommand{\DSoneH}{{\cal H}^{(\Delta S=1)}}

\title{Im$A_0$, Im$A_2$, and $\epsilon^\prime$ from quenched lattice QCD}

\author{T. Blum\address{RIKEN-BNL Research Center,
Brookhaven National Laboratory, Upton, NY 11973-5000, USA}\\
{RBC Collaboration}
\thanks{This work was done in collaboration with
N.~Christ, C.~Cristian, C.~Dawson, G.~Fleming,
X.~Liao, G.~Liu, S.~Ohta, A.~Soni, P.~Vranas, R.~Mawhinney,
M.~Wingate, L.~Wu, and
Y.~Zhestkov.  We thank RIKEN, Brookhaven National Laboratory and the
U.S.\ Department of Energy for providing the facilities essential for
this work.}}

\newcommand{\bea}{\begin{eqnarray}}
\newcommand{\eea}{\end{eqnarray}}

\begin{document}

\begin{abstract}
We present results for the imaginary parts of the $I=0$ and 2 
$K\to\pi\pi$ decay amplitudes, Im$A_{0,2}$, and the ratio of
$CP$ violation parameters, $\epsilon^\prime/\epsilon$. Our calculation
is done in the quenched approximation at $a^{-1}\approx 2$ GeV, lattice
size $16^3\times 32$, using domain wall fermions with $L_s=16$. We study the
three flavor case (charm is not an active flavor) 
and find $\epsilon^\prime/\epsilon$ is small and
slightly negative.
\vspace{1pc}
\end{abstract}
\maketitle

\section{Introduction}

In this talk we present results for the
$CP$ violating imaginary parts of the isospin zero and two
$K\to\pi\pi$ decay amplitudes, and
hence the theoretical value of $\epsilon^\prime$. We build on the
talks by Mawhinney and Cristian where details of the basic method
and simulations can be found. Cristian presented results for
the $CP$ conserving real parts of the amplitudes, and
it was found that the theoretical calculation of the ratio of the real parts
was in good agreement with the long-standing $\Delta I=1/2$ rule. Here we find
the imaginary parts, together with the real parts, yield a value of
$\epsilon^\prime/\epsilon=(-4.0\pm 2.3)\times 10^{-4}$ (statistical 
errors only) that is small and
slightly negative, in contrast to the recently determined 
values, $\sim(15-20)\times 10^{-4}$, from the NA48 and KTeV experiments
at CERN and Fermilab.
We refer the reader to $\cite{us}$ for complete details of our calculation.
The CP-PACS collaboration has also presented a very
similar calculation at this meeting\cite{CPPACS} and finds results for
$\epsilon^\prime/\epsilon$ that are compatible with ours.

Defining decay amplitudes as
\bea
A(K^0\to\pi\pi(I))&=&A_I\,e^{i \delta_I},\nonumber\\
A(\overline{K^0}\to\pi\pi(I))&=&-A_I^*\,e^{i \delta_I},
\eea
where $I=0$, 2 is the isospin of the
final state pions and $\delta_I$ the corresponding s-wave phase shift, 
it can be shown that 
\begin{equation}
  \frac{\epsilon^\prime}{\epsilon} = 
   \frac{i\,e^{-i(\delta_0-\delta_2)}}{ \sqrt{2}\, \epsilon }
        \frac{{\rm Re\,A}_2}{{\rm Re\,A}_0}
        \left( 
        \frac{{\rm Im\,A}_2}{{\rm Re\,A}_2} -
        \frac{{\rm Im\,A}_0}{{\rm Re\,A}_0}\right),\nonumber
\end{equation}
where we have neglected isospin breaking effects (these are expected to
decrease the value of $\epsilon^\prime$).
The factors outside of the parentheses are 
well known experimentally, so we quote
values of $\epsilon^\prime/\epsilon$ using these known values.
Thus, the calculation boils down to determining the difference of
ratios $P_I\equiv{\rm Im}A_I/{\rm Re}A_I$.

Im$A_{0}$ and  Im$A_{2}$ are given by 
${\rm Im}\,(e^{-i\,\delta_I}\times$ $\langle
\pi\pi(I)|{-i\DSoneH}|K^0\rangle)$
where the effective weak Hamiltonian, $\DSoneH$,
is a linear combination of four-quark operators with Wilson 
coefficients determined by the Standard Model. 
The imaginary part of $\DSoneH$ arises from the
$CP$ odd phase of the CKM mixing matrix.
The non-perturbative calculation of the
matrix elements of the four-quark operators forms the
fundamental core of our calculation (see the talk by Mawhinney). We use
chiral perturbation theory to extract the low energy constants of $\DSoneH$
from calculations of unphysical $K\to\pi$ and $K\to 0$ matrix elements.
This procedure directly yields the corresponding $K\to\pi\pi$ amplitudes to
lowest order in chiral perturbation theory \cite{Bernard:1985wf}
(${\cal O}(p^0,\,p^2)$ depending on
the particular operator). Some of the
${\cal O}(p^4)$ corrections to these relations 
have been calculated \cite{Bijnens:1985qt,Cirigliano:1999pv}, 
which we include in our analysis.
However, a complete next-to-leading order calculation using 
only the reduced matrix elements cannot be done 
since the relevant ${\cal O}(p^4)$
counter-terms in the $\Delta S=1$ chiral Lagrangian differ in the two
cases. It must therefore be argued that these higher order
effects are small in the
final answer. With the low energy constants and Wilson coefficients 
in hand, it is straightforward to calculate the physical decay
amplitudes.
%

\section{Results}
\begin{figure}[h]
\includegraphics[width=3in]{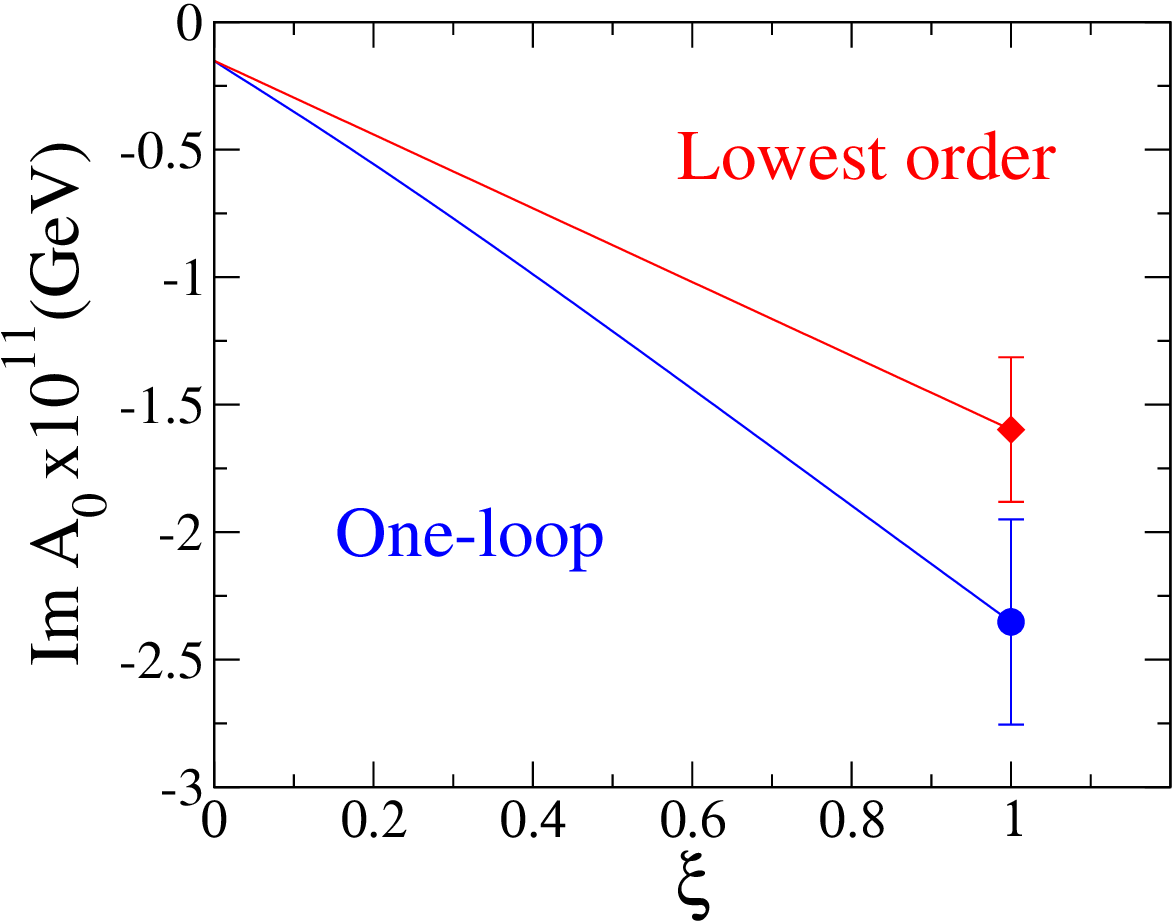}
\vskip -.25in
\caption{The imaginary part of $A_0$. 
One loop chiral log corrections are large at the
physical point, $\xi=1$. The errors are statistical only.} 
\label{imag0}
\end{figure}

In Fig.~\ref{imag0} we show Im$A_0$ as a function of a fictitious mass
parameter $\xi$ ($m_K^2,\,m_\pi^2\to \xi \times(m_K^2,\,m_\pi^2)$)
which we have introduced in order to study the
behavior of the ``physical amplitudes'' from the chiral limit ($\xi=0$)
to the actual 
physical point ($\xi=1$). In particular, our calculation should be
increasingly accurate as $\xi\to 0$. 
We show results using 
lowest order chiral perturbation theory as well as 
those with one-loop corrections. The amplitude is dominated by the 
QCD penguin operator $Q_{6}$ which vanishes in the chiral limit. 
It does not vanish at $\xi=0$ due to the small
contribution of the electroweak penguin operators $Q_{7,8}$. We note that
the one-loop chiral log corrections are significant (roughly 50\%).

\begin{figure}[h]
\includegraphics[width=3in]{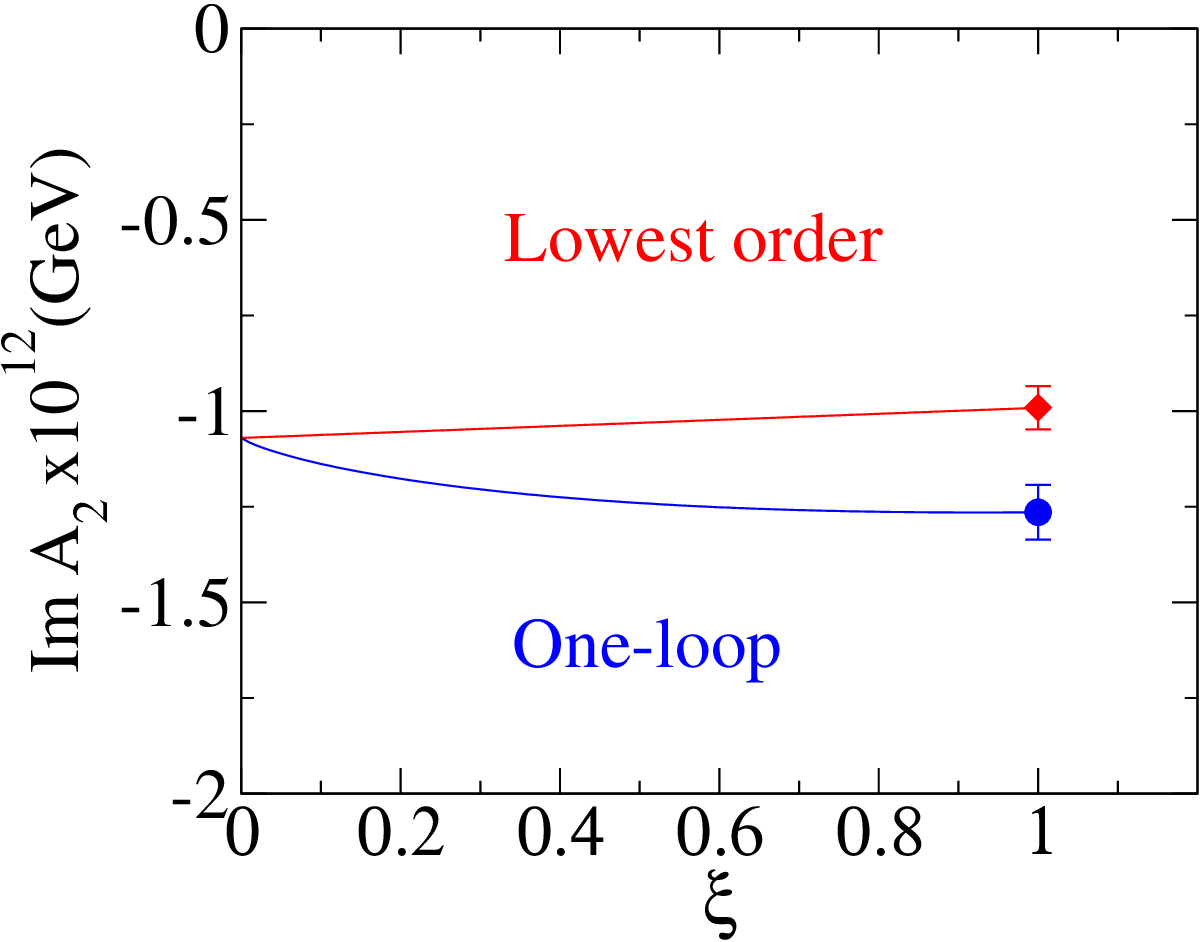}
\vskip -.25in
\caption{Same as Fig.~\ref{imag0} except for Im$A_2$.}
\label{imag2}
\vskip -.2in
\end{figure}

Similarly, in Fig.~\ref{imag2} we show Im$A_2$. Since Im$A_2$ is dominated by
$Q_8$ whose matrix element is a constant at lowest order, it is also
roughly constant over the whole range, and 
the one-loop corrections are not as large as for the $I=0$ channel.
It is interesting to note that at the physical point the imaginary parts
of the amplitudes also exhibit a ``$\Delta I=1/2$'' rule similar to the 
real parts.

\begin{figure}[h]
\includegraphics[width=3in]{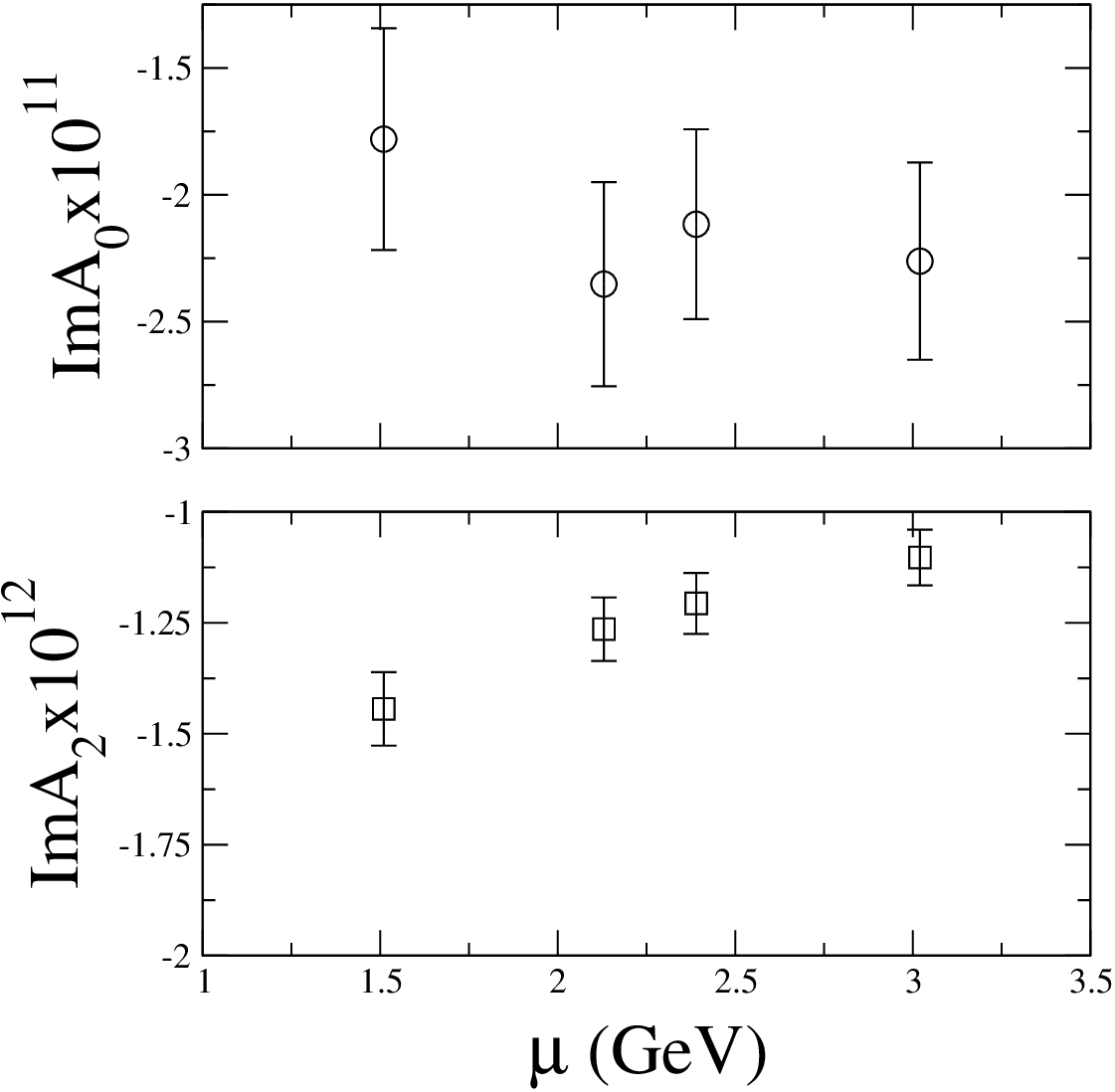}
\vskip -.25in
\caption{Residual renormalization 
scale dependence in the physical amplitudes (units are GeV).}
\label{res_dep}
\vskip -.25in
\end{figure}

The residual renormalization scale dependence in the physical amplitudes
is shown in Fig.~\ref{res_dep} and is most apparent for the $I=2$ case
where 
the statistical errors are smaller. In both cases, the largest deviation is at
the lower end.
      Had the Wilson coefficients been calculated to all
      orders in perturbation theory, had our matrix elements
      been evaluated in full QCD, and had we included
      subtractions of all operators with lower dimension (not just
      power divergent ones) this
      scale dependence would have canceled exactly in
      ${\cal H}^{\Delta S=1}$ since its matrix elements are
      physical observables.
As $\mu$ decreases non-perturbative effects in the operators
may arise which are not compensated in the Wilson coefficients. On the
other-hand, if $\mu$ becomes too large, lattice artifacts appear in the
renormalized operators. Thus we quote values for physical
quantities at $\mu=2.13$ GeV (of course, 
the scale is essentially determined by the inverse lattice
spacing, $a^{-1}=1.922$ GeV in this case).

\begin{figure}[b]
\includegraphics[width=3in]{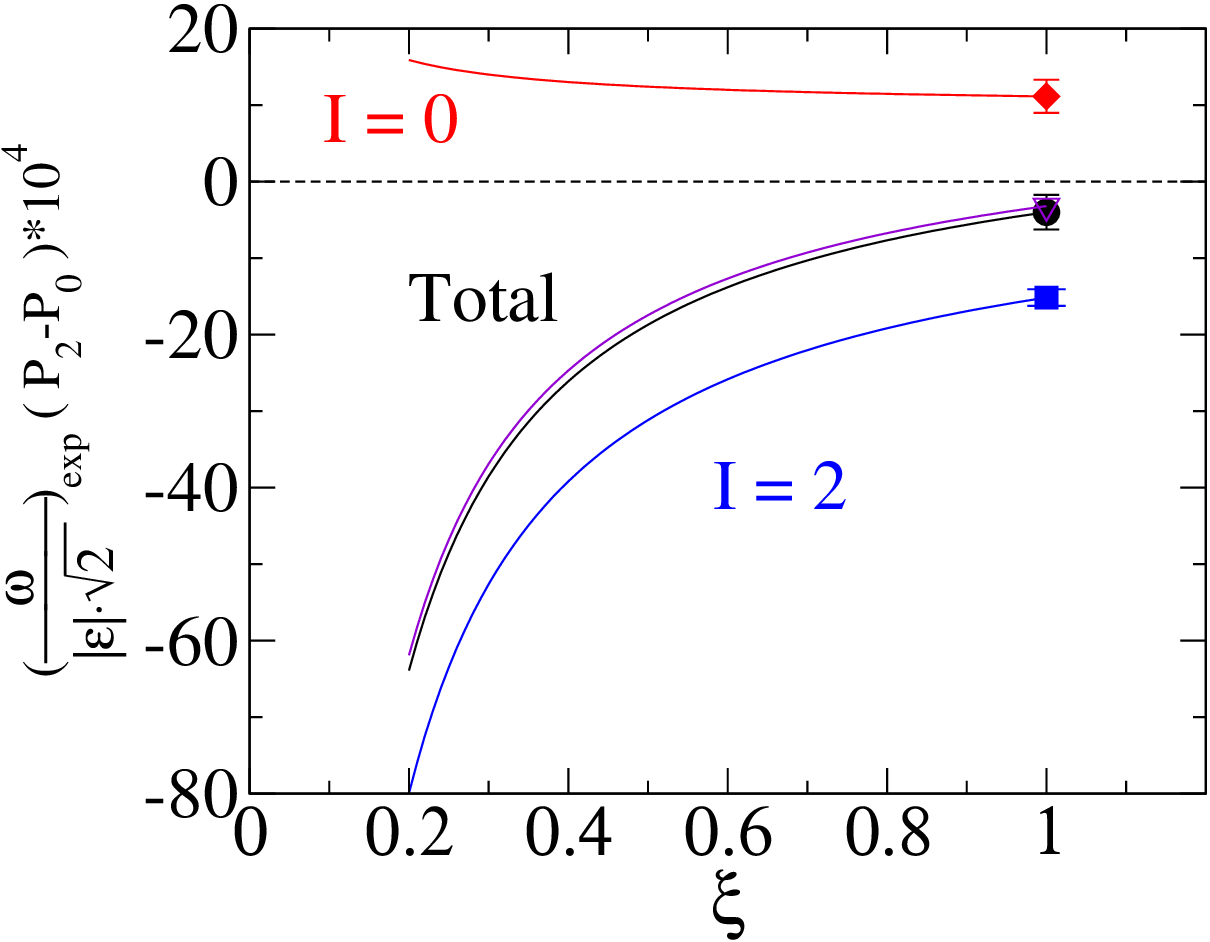}
\vskip -.25in
\caption{The ratio of $CP$ violation parameters $\epsilon^\prime/\epsilon$.
Totals are shown for lowest order (upper) 
and one-loop (lower) chiral perturbation theory.
The $I=0$ and 2 contributions are shown for the one-loop case only.
The experimental value is between 15 and 20.}
\label{epsp}
\end{figure}

Our results for $\epsilon^\prime/\epsilon$ are shown in Fig.~\ref{epsp}.
The value at the physical point is small and slightly negative whereas the
experimental value lies somewhere between 15 and 20 parts in $10^{-4}$.
Interestingly, the one-loop corrections that were significant in the
individual amplitudes do not play any role here. Just away
from the chiral limit, both the numerator and denominator of $P_0$ are 
dominated by (8,1) operators $Q_6$ and $Q_2$, respectively, so the
log correction (which is the same in both cases) cancels. 
Since the leading behavior of both is $\xi(m_K^2-m_\pi^2)$, $P_0$ behaves 
essentially like a constant (see Fig.~\ref{epsp}).
$P_2$ is dominated by $Q_8$ in the numerator and $Q_2$ in the denominator.
The corresponding one-loop
corrections are not the same; however, they are similar in magnitude, 
so the net effect is
not large. Since Re$A_2$ is dominated by (27,1) operators, it (almost) vanishes
as $\xi\to0$ whereas we saw Im$A_2$ was roughly
constant. So, as $\xi\to0$, $\epsilon^\prime/\epsilon$ becomes large and
negative (note that at $\xi=0$, $P_0=P_2$ and $\epsilon^\prime/\epsilon=0$).

\section{Summary and Outlook}
We have completed a first 
calculation of $\epsilon^\prime/\epsilon$
in the Standard Model, albeit with significant approximations including
quenching, use of perturbation theory below $m_{\rm charm}$,
and incomplete next-to-leading order calculations in 
chiral perturbation theory. Such approximations could
conspire to yield the current result which is in disagreement with
experiment, even though
it is difficult to point to a single approximation that may be
responsible. In any case, the above approximations are not insurmountable and
will be addressed in future calculations. 
     In particular, the results
     reported here are the first part of a two-part
     calculation.  The second part, not yet finished,
     uses an active charm quark and the four-quark effective
     theory.
The road ahead will be interesting, especially if we 
must turn away from the Standard Model.

\vskip -.5in

\end{document}